# Aligning long-term climate mitigation with enhanced methane action

Katsumasa Tanaka[1,2,*,] , Kushal Tibrewal[1,*], Philippe Ciais[1], Olivier Boucher[3]

[1] Laboratoire des Sciences du Climat et de l'Environnement (LSCE), IPSL, CEA-CNRS-UVSQ, Université Paris-Saclay, Gif-sur-Yvette, France

[2] Earth System Division, National Institute for Environmental Studies (NIES), Tsukuba, Japan

[3] Institut Pierre-Simon Laplace (IPSL), Sorbonne Université / CNRS, Paris, France

* These authors contributed equally to this work.

Corresponding author: katsumasa.tanaka@lsce.ipsl.fr

## Abstract

The Global Methane Pledge and other methane measures may potentially undermine $CO_2$ mitigation in certain countries, unless they are considered as additional to the existing Nationally Determined Contributions to strengthen overall greenhouse gas emission targets. Maintaining the progress on $CO_2$ mitigation in the revision of Nationally Determined Contributions after the first Global Stocktake, while pursuing the immediate benefits from methane mitigation, is necessary to address climate change in the long-term.

## Main text

The Global Methane Pledge (GMP)[1] and subsequent political development, such as the US-China Sunnylands Statement[2] prior to the 28th Conference of the Parties (COP28) in December 2023, provide the impetus for enhanced methane ($CH_4$) action as part of climate mitigation strategies[3]. The GMP calls for 30% $CH_4$ emission reductions by 2030 relative to 2020 levels globally. To date, GMP is joined by 155 countries, covering more than 50% of the global anthropogenic $CH_4$ emissions, although some major emitters, including China, India and Russia, are yet to join. Recent improved mapping of $CH_4$ ultra-emitters from the oil and gas sector[4], emerging evidence of $CH_4$ feedback in the Earth system[5] and the strong recent upward trend of atmospheric $CH_4$ concentrations[6] also support such an initiative.

While reducing $CH_4$ emissions is effective for limiting near-term warming and improving air quality and human health[7,8], reducing carbon dioxide ($CO_2$) emissions is essential for ultimately halting and reversing long-term warming toward the 1.5 °C level after temperature overshoot[9–11]. Parallel pursuit of these goals is necessary, taking into account the very different atmospheric properties of $CO_2$ and $CH_4$[12]. Mitigation priority of long-lived climate forcers (e.g., $CO_2$) over short-lived climate forcers (e.g., $CH_4$ and black carbon), or vice versa, has long been debated at the global[13–22], regional and sectoral[23–25] levels. The debate also mirrors the discussion of emission metrics since around 1990[26] as synthesized in the Intergovernmental Panel on Climate Change (IPCC) Sixth Assessment Report (AR6)[27]. The recent focus on $CH_4$ mitigation has injected a renewed sense of urgency into this debate.

Following the first Global Stocktake concluded at COP28, Parties to the Paris Agreement will revise their Nationally Determined Contributions (NDCs). NDCs have typically been expressed in terms of $CO_2$-equivalent ($CO_2$eq) emissions for the near-term (usually 2030), combining emission reductions of $CO_2$, $CH_4$ and other greenhouse gases (GHGs). Note that there are also NDCs that only consider $CO_2$ and that certain NDCs also include mid-century net-zero targets. Parties have approximately one year to revise their current NDCs and another year to include 2035 targets (Paragraphs 166 and 170 of the UNFCCC document[28], respectively). Although the Paris signatories are generally not explicit about how enhanced $CH_4$ targets will be treated in their NDCs, it is reasonable to assume that enhanced $CH_4$ targets will also be considered (or should be considered for consistency) as part of their near-term GHG targets in NDCs. Some countries may find it easier to mitigate $CH_4$ than $CO_2$ by targeting cost-effective measures, such as reducing $CH_4$ leakage in the oil and gas sector[4,29]. While strengthening $CH_4$ mitigation should be welcomed, we argue that it is important to ensure that progress on $CO_2$ mitigation is not compromised at the expense of enhanced $CH_4$ mitigation.

Here, we show that if countries' $CH_4$ targets are enhanced through GMP and introduced into their NDCs but internally compensate their $CO_2$ targets stated or implied in their NDCs (Methods), such target implementations can result in a significant setback for long-term climate mitigation. To demonstrate this point, we use illustrative scenarios below.

- Scenario A (NDCs and net-zeros): This reference scenario accounts for latest national climate pledges, such as 2030 emission targets and mid-century net-zero targets in NDCs[30] and Long-Term Strategies (LTS)[31] by the end of COP28 (cut-off date: 13 December 2023). Enhanced $CH_4$ targets are not considered, as they are actually not yet reflected in many NDCs.

- Scenario B (NDCs, net-zeros, and enhanced $CH_4$ targets *without compensating* $CO_2$ targets): If a country's current $CH_4$ target stated or implied in its NDC (unconditional target only; Methods) is smaller than the enhanced $CH_4$ target, the difference will be added to its current $CH_4$ target and consequently its total 2030 GHG target in the NDC (i.e., $CO_2$eq emissions using the 100-year Global Warming Potential (GWP100)).

- Scenario C (NDCs, net-zeros, and enhanced $CH_4$ targets *with compensating* $CO_2$ targets to keep the respective total GHG targets in NDCs unchanged): For countries meeting the criteria above, the difference in the $CH_4$ target levels will be subtracted from the $CO_2$ target level in the NDC to maintain the same total 2030 GHG targets.

For Scenarios B and C, we consider two different penetration levels of GMP: i) current participating countries (Scenarios B1 and C1) and ii) all countries (Scenarios B2 and C2). Our scenarios are based on a simple and transparent method[32] to interpolate emissions between targets and extrapolate emissions to 2100 (Methods), which produces comparable results with previous studies[33,34]. Note that we focus on the trade-off between $CO_2$ and $CH_4$ targets without explicitly accounting for measures that mitigate both gases simultaneously.

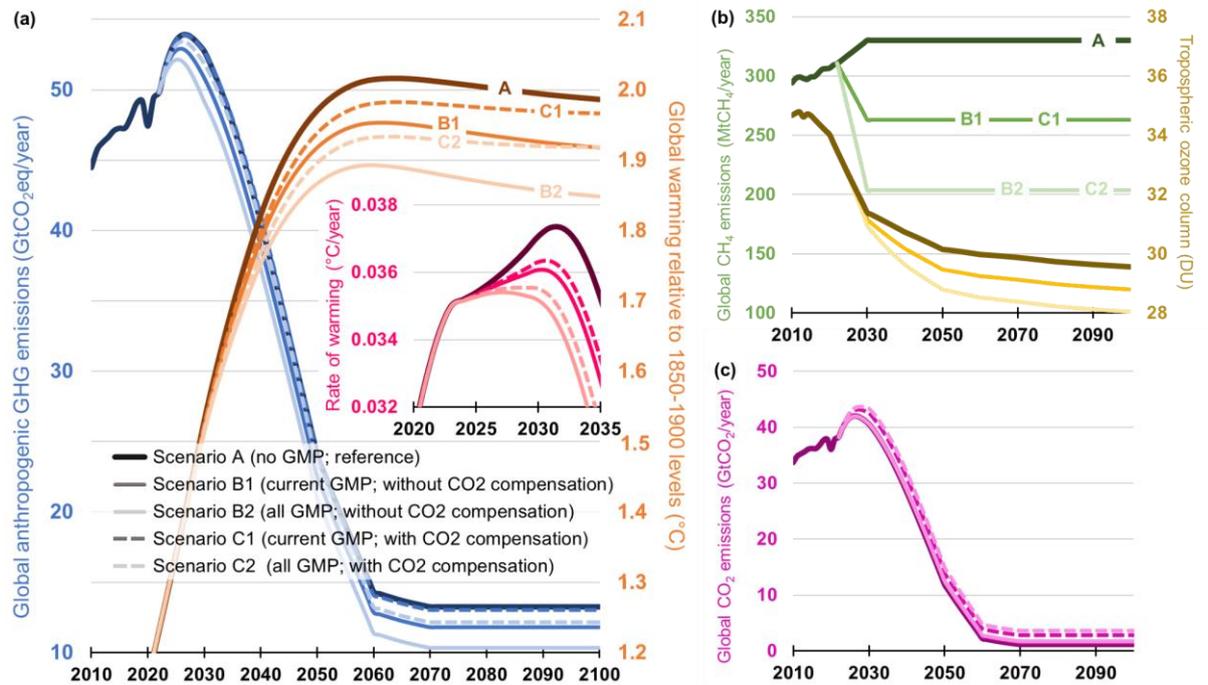

**Fig. 1 | Potential impacts of enhanced $CH_4$ mitigation on the global anthropogenic $CO_2$, $CH_4$ and GHG emission, tropospheric ozone concentration and temperature pathways with or without compensating $CO_2$ targets.** Scenario A is the reference scenario without GMP. Scenarios B1 and B2 consider GMP for current participating countries and all countries, respectively, *without* compensating $CO_2$ targets. Likewise, Scenarios C1 and C2 also consider GMP but *with* compensating $CO_2$ targets. Scenario labels are given for selected lines only. Blue and orange lines in panel **a** are the global total anthropogenic GHG emissions (i.e., the sum of anthropogenic $CO_2$, $CH_4$ and nitrous oxide ($N_2O$) emissions using IPCC AR5 GWP100 values for GHG aggregation, excluding $CO_2$ emissions from the Land Use, Land-Use Change and Forestry (LULUCF) sector) (GtCO$_2$eq/year, left scale) and the global mean surface air temperature increase relative to 1850-1900 levels (°C, right scale), respectively. Panel **a** also shows the rate of near-term warming (°C/year) in red lines in insert. Green and yellow lines in panel **b** are the global total anthropogenic $CH_4$ emissions (MtCH$_4$/year) and tropospheric ozone column (DU), respectively. Pink lines in panel **c** are the global total anthropogenic $CO_2$ emissions excluding LULUCF (GtCO$_2$/year).

Our calculations using the reduced-complexity climate model ACC2[9,35,36] show that the peak warming, as well as the rate of near-term warming, can be reduced with enhanced $CH_4$ mitigation, along with co-benefits for air quality (reduced tropospheric ozone column) (Fig. 1). However, the benefit of lower warming does not persist in the long run if enhanced $CH_4$ targets compensate for existing $CO_2$ targets (Scenarios C1 against B1). The end-of-century warming will still become lower in Scenario C1 (by -0.020 °C) than the reference level in Scenario A (1.987 °C), but such a gain in warming for Scenario C1 is less than half of the gain in Scenario B1 (by -0.069 °C) and will eventually vanish after 2100. To our knowledge, this is the first study based on actual climate pledges to confirm previous findings[19,20] conducted in more idealized settings without such pledges. If we assume a full penetration of GMP, the results become more pronounced (Scenarios B2 and C2).

At the country level, the long-term temperature outcome is largely determined by the mid-century target (all scenarios in Fig. 2). Enhanced $CH_4$ targets consistently improve the long-term temperature outcome (Scenarios B against A); however, such a gain in warming can be diminished in different ways by $CO_2$ target compensations (Scenarios C against B).

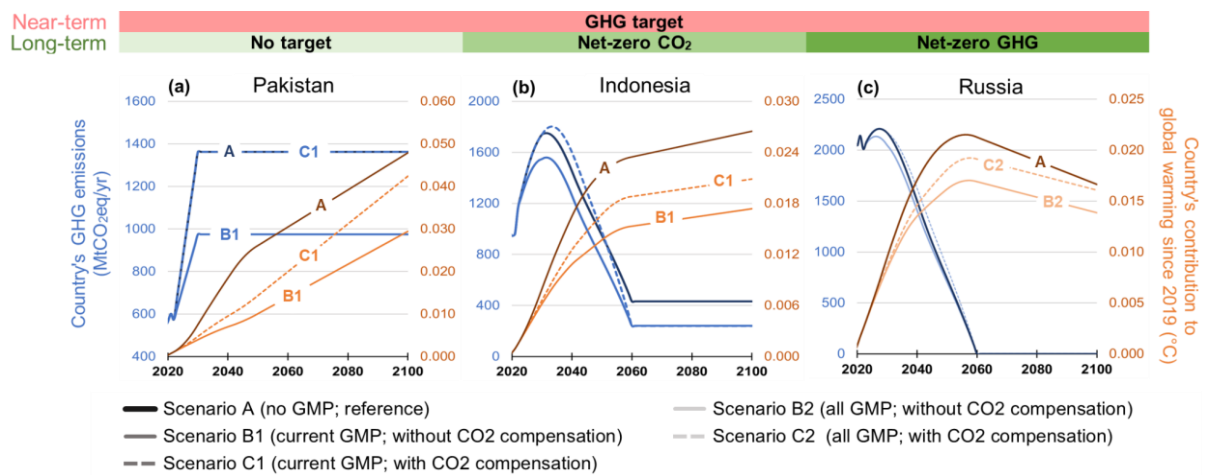

**Fig. 2 | Potential impacts of enhanced $CH_4$ mitigation on the anthropogenic GHG emission pathways of selected example countries and the implications for their future respective contributions to global warming with or without compensating $CO_2$ targets.** The figure shows

different ways in which enhanced $CH_4$ mitigation can affect countries' GHG emissions and global temperatures under different long-term targets. The panels show the total anthropogenic GHG emissions of each country (i.e., the sum of anthropogenic $CO_2$, $CH_4$ and $N_2O$ emissions using IPCC AR5 GWP100 values for GHG aggregation, excluding LULUCF) (MtCO$_2$eq/year) and its contribution to the global mean surface air temperature increase since 2019 (°C). Scenario labels are given for selected lines only. Scenarios B1 and C1 are shown for Pakistan and Indonesia as they are participants in GMP. Scenarios B2 and C2 are for Russia as it is not a signatory to GMP. These countries are selected for illustrative purposes. It is important to note that our future projections are made under simplifying assumptions and should be viewed as benchmarks to analyze different target implementations, not as realistic forecasts.

- Countries with no net-zero target: The temperature continues to rise through 2100 because $CO_2$ and $CH_4$ emissions after 2030 are assumed constant with no further targets (Scenario A). A near-term increase in $CO_2$ emissions as a consequence of the target compensation also persists through 2100, largely offsetting the earlier gain in warming (Scenarios C1 against B1). Such results are particularly relevant for countries that have limited $CH_4$ targets but with abundant low-cost $CH_4$ mitigation opportunities from the oil and gas sector, such as Turkmenistan, Uzbekistan and the Eastern Mediterranean and Middle East (EMME) countries[4,29] (Supplementary Fig. 1).

- Countries with a net-zero $CO_2$ target: The temperature continues to rise moderately after the net-zero year because of continuing non-$CO_2$ emissions (Scenario A). An increase in $CO_2$ emissions from the target compensation diminishes by the net-zero year. The earlier gain in warming is maintained thereafter (Scenarios C1 against B1). However, to avoid compromising any gain in warming, it is advisable for countries, such as China and India, not to substitute planned $CO_2$ mitigation with any future $CH_4$ mitigation.

- Countries with a net-zero GHG target: The temperature declines after the net-zero year because of the net negative $CO_2$ emissions that offset the residual non-$CO_2$ emissions to achieve the GHG net-zero target as determined by GWP100[9,37] (Scenario A). Enhanced $CH_4$ mitigation could lower future decarbonisation rates, as it implies lower $CH_4$ emissions from 2030 onwards, requiring less negative $CO_2$ emissions to achieve the net-zero GHG target (Scenarios C2 vs. B2). It is thus recommended for countries, such as Australia, the EU states and the US, to maintain their planned decarbonization rates when more stringent $CH_4$ mitigation is considered.

The different near- and long-term temperature effects discussed here reflect the classic problem associated with GWP100, the emission metric stipulated in the Paris Rulebook[38] and commonly used elsewhere to convert GHG emissions on the common scale of $CO_2$ (e.g., 1 mass unit of $CH_4$ emissions is considered equal to 28 units of $CO_2$ emissions as GWP100 is 28 for $CH_4$ in IPCC AR5). GWP100 is not designed to maintain equivalence in the temperature trajectory upon the emission conversion[39–41] (but in the radiative forcing averaged over 100 years). An approach to ensuring temperature equivalence has been proposed and is being debated[42–45]; however, there are practical barriers to implementing this approach for the use of emission offsets[46], which would also apply to target compensations. As the same GHG pathways with different gas mixes can lead to very different temperature outcomes, clarification of how the $CH_4$ contribution is accounted for, which is lacking in many NDCs, could help avoid ambiguity in monitoring progress towards the 1.5 °C target[12].

In conclusion, enhanced $CH_4$ targets should be integrated into the respective NDCs for consistency. We argue that in doing so, countries should strengthen their NDCs jointly with enhanced $CH_4$ targets (if applicable) without relaxing their $CO_2$ targets. This is particularly important for countries with abundant cost-effective $CH_4$ mitigation opportunities in the oil and gas sector but not yet part of GMP, to ensure that enhanced methane action goes hand in hand with long-term climate mitigation.

**Methods**

The emission projections presented in this study are derived from our recently developed framework, as detailed in ref[32]. The framework aims to provide emission and temperature projections until 2100, based on the latest national climate pledges, and to monitor progress towards the Paris Agreement targets in a simple and transparent way accessible to various users. In our future emission projections, the most recent emission levels, near-term targets, and long-term targets (if available) are interpolated using a simple mathematical function[47]. If there is no further target, we assume that emissions will be constant beyond that point (i.e., Constant Emissions approach[48]). For example, if a country does not have a net-zero target in mid-century, we assume that the country's emissions will remain constant after 2030 through 2100. For a country with a net-zero target, the emissions are assumed to be constant thereafter at the levels of the net-zero target year. When the net-zero target is given for GHGs, which implies net negative $CO_2$ emissions to compensate for residual non-$CO_2$ emissions, the net negative $CO_2$ emissions are kept until 2100. If the type of net-zero target is not specified in the NDC, a conservative assumption is made that the net-zero target applies only to $CO_2$. Our analysis focuses on the three main GHGs ($CO_2$, $CH_4$, and $N_2O$). We assume that emissions of other GHGs and air pollutants follow SSP1-1.9[49].

If a country's $CH_4$ mitigation target is not specifically given in its NDC (applicable only to countries with NDCs given for GHGs), we infer the $CH_4$ mitigation target level from the 2030 GHG target level under the following assumptions: i) The proportion of $CO_2$ and $CH_4$ emissions (in terms of $CO_2$eq emissions) will be constant during 2020-2030. ii) The magnitude of $N_2O$ emissions will be constant over this period. iii) These three gases make up the bulk of GHGs covered in the NDCs. If an NDC only concerns $CO_2$ (e.g., India and China (subject to interpretation[25])), no compensations with $CO_2$ will be considered.

According to the official GMP website[1], the signatories agree to contribute to a collective effort to reduce global $CH_4$ emissions by at least 30% by 2030 compared to 2020 levels. This may be interpreted to mean that participating countries will collectively aim for a 30% reduction in global $CH_4$ emissions, including emissions from non-participating countries. However, our study does not subscribe to this interpretation. We consider only 30% $CH_4$ reductions for countries participating to GMP, regardless of the level of $CH_4$ emission reduction by non-participating countries.

The temperature calculations are performed using a reduced-complexity climate model ACC2[9,35,36]. The model consists of four modules: carbon cycle, atmospheric chemistry, physical climate, and mitigation. The mitigation module allows the model to compute least-cost emission pathways for a given climate target through optimization, but this module is not used in this study and ACC2 is used only as a simulator. ACC2 takes into account major feedback processes in the Earth system, such as $CO_2$ fertilization, saturation of ocean $CO_2$ uptake with increasing $CO_2$ concentrations, climate-carbon cycle feedback, OH chemistry to calculate tropospheric ozone concentrations from $CH_4$ concentrations and pollutant emissions. The equilibrium climate sensitivity is assumed to be 3 °C, the best estimate of the IPCC AR6, within its range of 2.5 °C to 4 °C, with other uncertain parameters optimized in a historical inversion setup[50].

**Method References**

**Data and materials availability** All data needed to evaluate the conclusions in the paper are present in the paper and/or the Supplementary Information and are available on Zenodo with doi.org/XXX.

**Acknowledgments** This work was conducted as part of the Achieving the Paris Agreement Temperature Targets after Overshoot (PRATO) Project under the Make Our Planet Great Again (MOPGA) Program and benefited from State assistance managed by the National Research Agency in France under the Programme d'Investissements d'Avenir under the reference ANR-19-MPGA-0008. This project has received funding from the European Union's Horizon Europe research and innovation program under Grant Agreement N° 101071247 (Edu4Climate – European Higher Education Institutions Network for Climate and Atmospheric Sciences). K. Tibrewal is supported by the European Union's Horizon 2020 research and innovation program under Grant Agreement N° 856612 (EMME-CARE – Climate change, atmospheric research centre in Eastern Mediterranean and Middle East Region). We further acknowledge the European Union's Horizon Europe research and innovation program under Grant Agreement N° 101056939 (RESCUE – Response of the Earth System to overshoot, Climate neUtrality and negative Emissions), Grant Agreement N° 101081193 (OptimESM – Optimal

High Resolution Earth System Models for Exploring Future Climate Changes) and Grant Agreement N° 820829 (CONSTRAIN – Constraining uncertainty of multi-decadal climate projections).

**Data and materials availability** K.Ta and K.Ti conceived this study. K.Ta led this study. K.Ta, and K.Ti designed the experiment. K.Ti performed the analysis. K.Ta and K.Ti. analyzed the results. K.Ta and K.Ti generated the figures. K.Ta. drafted the manuscript, with contributions from K.Ti., P.C., and O.B.

**Competing interests** The authors declare that they have no competing interests.

**Supplementary Information** Supplementary information for this article is available.

# Aligning long-term climate mitigation with enhanced methane action

# – Supplementary Information –

Katsumasa Tanaka[1,2,*,], Kushal Tibrewal[1,*], Philippe Ciais[1], Olivier Boucher[3]

[1] Laboratoire des Sciences du Climat et de l'Environnement (LSCE), IPSL, CEA-CNRS-UVSQ, Université Paris-Saclay, Gif-sur-Yvette, France

[2] Earth System Division, National Institute for Environmental Studies (NIES), Tsukuba, Japan

[3] Institut Pierre-Simon Laplace (IPSL), Sorbonne Université / CNRS, Paris, France

* These authors contributed equally to this work.

Corresponding author: katsumasa.tanaka@lsce.ipsl.fr

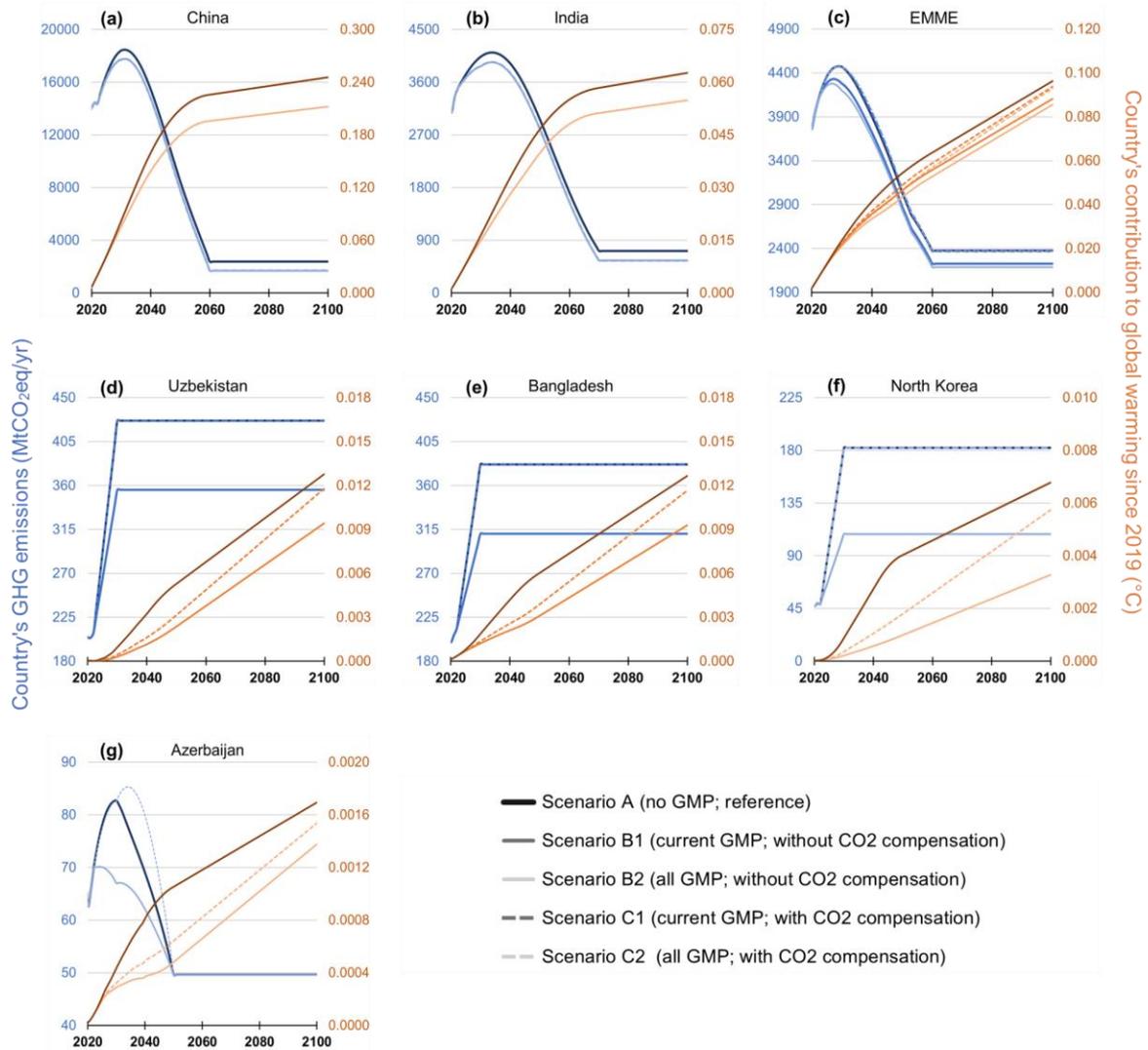

**Supplementary Fig. 1 | Potential impacts of enhanced CH$_4$ mitigation on the anthropogenic GHG emission pathways of selected example countries and the implications for their future respective contributions to global warming with or without compensating CO$_2$ targets.** This figure supplement Fig. 2 of the main paper. The figure shows different ways in which enhanced CH$_4$ mitigation can affect countries' GHG emissions and global temperatures under different long-term targets. The panels show the total anthropogenic GHG emissions of each country (i.e., the sum of anthropogenic CO$_2$, CH$_4$ and N$_2$O emissions using IPCC AR5 GWP100 values for GHG aggregation, excluding LULUCF) (MtCO$_2$eq/year) and its contribution to the global mean surface air temperature increase since 2019 (°C). These countries are selected for illustrative purposes. It is important to note that our future projections are made under simplifying assumptions and should be viewed as benchmarks to analyze different target implementations, not as realistic forecasts.